%% file: main.tex
\documentclass[letterpaper]{article} 
\usepackage{aaai25}  
\usepackage{times}  
\usepackage{helvet}  
\usepackage{courier}  
\usepackage[hyphens]{url}  
\usepackage{graphicx} 
\usepackage{natbib}  
\usepackage{caption} 
\usepackage{subfigure}
\usepackage{algorithm}
\usepackage{algorithmic}

\usepackage{newfloat}
\usepackage{enumitem}
\usepackage{listings}
\usepackage{multirow}
\usepackage[utf8]{inputenc}
\usepackage{ucs}
\DeclareCaptionStyle{ruled}{labelfont=normalfont,labelsep=colon,strut=off} 
\floatstyle{ruled}
\newfloat{listing}{tb}{lst}{}
\floatname{listing}{Listing}
\title{Enduring Disparities in the Workplace: A Pilot Study in the AI Community}
\author{
    Yunusa Simpa Abdulsalam\textsuperscript{\rm 2}$^\dag$
, 
    Siobhan Mackenzie Hall\textsuperscript{\rm 5}, 
    Ana Quintero-Ossa\textsuperscript{\rm 4}, 
    William Agnew\textsuperscript{\rm 3}, 
    Carla Muntean\textsuperscript{\rm 1}\thanks{Work performed while volunteering for the organization},
    Sarah Tan\textsuperscript{\rm 1}, 
    Ashley Heady\textsuperscript{\rm 1}$^*$,
    Savannah Thais\textsuperscript{\rm 1}, 
    Jessica Schrouff\textsuperscript{\rm 1}\thanks{Corresponding author: jessicaschrouff@wimlworkshop.org}
}

\affiliations{
    \textsuperscript{\rm 1}Women in Machine Learning,
    \textsuperscript{\rm 2}Black in AI,
    \textsuperscript{\rm 3}Queer in AI,
    \textsuperscript{\rm 4}LatinX in AI,
    \textsuperscript{\rm 5}Deep Learning Indaba

}

\makeatletter
\usepackage{fancyhdr}  
\def\copyright@text{\footnotesize}

\begin{document}

\maketitle
%

\begin{abstract}
    \input{sections/00_abstract}
\end{abstract}

%

\input{sections/01_introduction}
\input{sections/02_related_works}

\input{sections/03_methods}

\input{sections/04_results}

\input{sections/05_discussion_conclusion}

\input{sections/98_acknowledgements}
\bibliography{main}
\bibliographystyle{aaai25}

\section*{Appendix}
\input{sections/99_appendix}

\end{document}

%% file: sections/00_abstract.tex
In efforts toward achieving responsible artificial intelligence (AI), fostering a culture of workplace transparency, diversity, and inclusion can breed innovation, trust, and employee contentment. In AI and Machine Learning (ML), such environments correlate with higher standards of responsible development. Without transparency, disparities, microaggressions and misconduct will remain unaddressed, undermining the very structural inequities responsible AI aims to mitigate. While prior work investigates workplace transparency and disparities in broad domains (e.g. science and technology, law) for specific demographic subgroups, it lacks in-depth and intersectional conclusions and a focus on the AI/ML community. To address this, we conducted a pilot survey of 1260 AI/ML professionals both in industry and academia across different axes, probing aspects such as belonging, performance, workplace Diversity, Equity and Inclusion (DEI) initiatives, accessibility, performance and compensation, microaggressions, misconduct, growth, and well-being. Results indicate enduring disparities in workplace experiences for underrepresented and/or marginalized subgroups. In particular, we highlight that accessibility remains an important challenge for a positive work environment and that disabled employees have a worse workplace experience than their non-disabled colleagues. We further surface disparities for intersectional groups and discuss how the implementation of DEI initiatives may differ from their perceived impact on the workplace. This study is a first step towards increasing transparency and informing AI/ML practitioners and organizations with empirical results. We aim to foster equitable decision-making in the design and evaluation of organizational policies and provide data that may empower professionals to make more informed choices of prospective workplaces.

%% file: sections/01_introduction.tex
\section{Introduction} \label{sec:intro}
Inequality, discrimination, and bias remain a constant reality for many in the AI community. Despite evidence that diverse and inclusive teams lead to better outcomes \cite{Page2007, Hunt2015}, microaggressions -- the subtle discrimination and biases marginalized groups experience throughout their lives~\cite{sue2007racial} -- and a poor sense of belonging and well-being remain pervasive in both academia and industry workplaces ~\cite{Gassam2019DeloittesNS, NEA2021Examples, fattoracci2023need, walton2019belonging, woodruffe2016countering, corlett2023manifestations, knekta2020evaluation, Nadal2014, corlett2023manifestations}. Moreover, these affect certain demographics disproportionately~\cite{blanck2021diversity, dopico2023queering}, resulting in low retention and underrepresentation of these groups in scientific fields~\cite{NSF2021, Kozlowski2022}. In AI/ML, for example, women make up only 22\%~\cite{InterfaceAI, HowardIsbell} of the workforce; Microsoft, Google, and Facebook are reported to have less than 5\% Black employees between 2018 and 2020~\cite{West2019, Brown2022, HowardIsbell}, and Amazon reported a 2020 corporate workforce comprising 7.2\% Black people, 7.5\% Latinx people, and 0.5\% Native American people~\cite{Amazon2020Workforce}. Workers from underrepresented groups have also been reported to be leaving the AI field~\cite{Brown2022} or experiencing glass ceilings~\cite{EEOC2024} in their careers, while facing additional challenges in entrepreneurship and startup creation, including limited access to funding, mentorship, and networks that are critical for success~\cite{RamosTorres2024}.

In the field of Responsible AI, studies have demonstrated that non-representative workforces result in poorer quality of services and models that are less able to capture diverse perspectives ~\cite{HowardIsbell, West2019, werder2024empower, sorensen2024roadmap, papagiannidis2025responsible}. Examples include gender occupational biases in language and vision models~\cite{hall2023visogender, kirk2021bias}, intersectional biases in commercial facial analysis systems \cite{Buolamwini2018} and poor AI model quality of service for individuals living with disabilities~\cite{massiceti2024explaining, wen2024find}. 

Internet-scraped datasets used to train AI models capture disproportionately harmful and toxic representations of underrepresented groups~\cite{birhane2021multimodal}, and commonly used filtering methods further exclude already underrepresented perspectives, such as LGBTQ+ people, older women and younger men, and non-Western perspectives~\cite{hong2024s}. Downstream applications that leverage AI can also inflict significant harm, including but not limited to privacy violations, and censoring and downranking queer content~\cite{queerinai2023queer}. 

This paper studies the perceived workplace experience of underrepresented groups in the AI/ML community, and questions whether enough has been done so far to ensure that AI research and development teams are diverse. Unlike prior research that discusses the representation of different demographic subgroups in the field~\cite{InterfaceAI, WEF2025, DeloitteWomenInAI2021}, we focus on the work environment and how it is experienced by AI/ML professionals in different demographic subgroups. We surveyed 1260 AI/ML professionals on their workplace experiences, yielding 796 complete survey responses. Our survey contained detailed questions on work environments, particularly on aspects such as belonging, performance, workplace DEI initiatives, well-being, and microaggressions. This pilot study is, to the best of our knowledge, the first to extensively survey AI professionals across roles, locations, companies/universities, and seniority levels. As such, these results provide valuable insights into not only the current state of the AI/ML community but also a strong foundation for future comparative studies. The main conclusions of our paper are as follows: 
\begin{itemize}
    \item There are large disparities in workplace experiences, resulting in unfavorable outcomes for underrepresented and/or marginalized groups.
    \item Accessibility and the inclusion of employees with a disability or chronic condition is one of the main challenges for employers.
    \item Reporting and handling misconduct in the workplace remains an ongoing issue.
    \item Patterns of microaggressions in AI/ML workplace reflect those of the broader society.
    \item Employers and members of majority groups can do more to encourage transparency in the workplace.
\end{itemize}

%% file: sections/02_related_works.tex
\section{Related Work} \label{sec:rel_works}

Despite the growing body of work examining algorithmic outcomes, dataset biases, and systemic inequalities embedded in technological systems~\cite{Buolamwini2018, birhane2021multimodal, hall2023visogender}, there remains a dearth of literature focusing on the lived experiences of those within the AI/ML workforce, particularly individuals from underrepresented demographic groups, or on how workforce inequalities may reinforce the very biases that fairness interventions seek to address. 

Global workforce studies have been conducted to infer the gender distribution in the global workforce. Data such as public professional profiles~\cite{InterfaceAI}, and aggregate industry and government data~\cite{WEF2025, DeloitteWomenInAI2021} have been analyzed to reveal that the AI workforce is composed of only 22\% women (with only 15\% women in at the senior executive level) and that Black and Hispanic professionals remain below 10\% in major tech companies. In the US, McKinsey \& Company has aggregated proprietary ~\cite{Hunt2015}, publicly available~\cite{Hunt2018DeliveringMcKinsey} and survey data~\cite{Krivkovich2024WomenMcKinsey} to understand the relationship between racial, ethnic and gender diversity, and a company’s financial results. The survey studies have been conducted annually since 2014, including 1000 companies in the US, with 480 000 women surveyed to investigate workplace experiences such as handling microaggressions and career progression prospects. A survey study by Deloitte ~\cite{Gassam2019DeloittesNS} was conducted in the US across multiple professions, focusing on microaggressions. Company diversity reports, including those from Amazon~\cite{Amazon2020Workforce} and analysis by~\citeauthor{West2019} and~\citeauthor{Brown2022}, rely on self-reported employment data to track representation by gender, race, and ethnicity. These highlight the limited presence of underrepresented groups in technical and leadership roles. Multiple organizations~\cite{holistic2025microaggressions, Gassam2019DeloittesNS, csizmadia2020uconn} developed survey questionnaires to evaluate the prevalence and experience of microaggressions.  The surveys in~\citet{holistic2025microaggressions} and~\citeauthor{csizmadia2020uconn} were limited to one institution and one behavior (sense of belonging and microaggressions respectively). Studies have also been conducted in academia, in the US, for example \citeauthor{knekta2020evaluation} validated a questionnaire on measuring a sense of belonging. Their target group was biology students at a single higher education institution and was motivated by the accepted idea that a sense of belonging increases student retention and achievement.    

While broader discussions on DEI initiatives in the workplace exist~\cite{West2019, Brown2022, papagiannidis2025responsible}, to the best of our knowledge no study has systematically explored how these issues manifest for workers at different intersections of identity within the AI/ML field. Our survey is the first of its kind to consider multiple dimensions across the work environment for better transparency, in efforts to achieve responsible AI.

%% file: sections/03_methods.tex
\section{Methods} \label{sec:methods}
This section details the survey creation and data analysis. We first describe the ethical considerations for this study and define different terms used throughout this paper.

\subsection{Ethical reviews and informed consent} \label{met:ethical}
Our study was approved by the Institutional Review Board (IRB) of Columbia University (ref: AAAU9608). All research procedures were conducted in accordance with standard ethical considerations \cite{wma2013helsinki} and the principles governing research of human subjects. Participants’ decision to take part in the study was voluntary and they could refuse to answer any questions. To ensure participant privacy, all data was collected anonymously and all analyses report aggregated data (n$\geq$30 across Employers, n$\geq$10 within Employers) without any identifiers. Prior to participation, all participants were provided with a detailed, unsigned consent form for a web-based online survey that outlined the purpose of the study, potential risks and immediate benefits, and privacy/confidentiality measures. We further ensured that no email addresses or IP addresses were captured when conducting the survey. The privacy and confidentiality of participants were ensured through encrypted cloud data storage, with access limited to authorized personnel on the IRB-approved list. 

\subsection{Nomenclature} \label{meth:nomen}
Below, we define different terms that are used throughout this work.

\noindent\textbf{AI Affinity groups}: organizations or groups of individuals that identify with an underrepresented group in the AI community. 

\noindent\textbf{Diversity, Equity, Inclusion (DEI)}: initiatives, such as those focusing on recruitment practices, workplace culture, team dynamics, career progression~\cite{ifcDigital2Equal, ifcSheWinsAfrica, ifcDEI}, which are put forward by an employer to counter historical imbalances, injustices, marginalisation, or stereotyping~\cite{roberson2006disentangling}.

\noindent\textbf{Employee}: the person who consented to fill in our survey. Their role must be related to researching or applying artificial intelligence/machine learning techniques. We refer to each Employee as a survey respondent or participant.

\noindent\textbf{Employer}: The company, university, institute, or any entity that acts as the Employer of the Employee. If the Employee had multiple Employers, they were asked to fill in the survey once per Employer and/or select only one Employer. Each participant was requested to fill in the survey for a current or recent ($<$1 year at time of response) employer.

\noindent\textbf{Minority group}: We define minority groups as individuals who have experienced historical marginalization, underrepresentation, or systemic disadvantages in various aspects of the workplace \cite{zhang2024counts, rupert2010commitment, neikirk2023recognizing}. In section \ref{sec:results}, we present the representation of demographic groups (\ref{subsec:rep-demo}), and highlight the minority groups identified in this study. 

\noindent\textbf{Majority group}: We define majority groups as individuals who have privileged or dominant positions in the society due to either social status, ethnicity, demographic characterization, and institutional/organizational power \cite{chi2021reconfiguring, radford2018conceptualizing, atewologun2008intersecting}. More details in representation of demographic subgroups (\ref{subsec:rep-demo}) in section \ref{sec:results}.

\noindent\textbf{Item}: A single question within a category.

\noindent\textbf{Question / Category / Construct}: These include multiple items aimed at probing the same variable.

\subsection{Survey design} \label{met:survey_design}
\input{tables/questions_table}
Our anonymous survey was designed by members of affinity groups with the help of researchers and sociologists specializing in investigating identity gaps in science. To effectively capture the differences between diverse subgroups, we categorized survey questions to align with key aspects that influence workplace transparency, such as belonging, performance, DEI initiatives, and microaggressions. Our selected categories focus on the dimensions most closely tied to uncovering systematic barriers that hinder workplace transparency. We provide the full details of these questions in Table \ref{table:survey_questions}. Participants were asked to rate questions on a 4-point Likert scale~\cite{Likert1932}, ranging from strongly agree to strongly disagree, and questions referring to frequencies from never to multiple times a day, structured as a 5-point Likert scale. Participants had the option to skip questions or answer ‘NA/I don’t know’. Some questions were conditioned on prior responses, e.g., when asking for more information about an Employer’s accessibility measures, or after requesting specific consent to provide details about experienced micro-aggressions. The survey was designed such that all questions were positively framed. For instance, this item in the Belonging question states “My perspective is valued, even if it is different from others.” Strongly agreeing with the statements within the survey indicated a positive and supportive work environment. To support this positive framing, the options were presented from left to right, from the most positive to the most negative answer. To avoid undecidedness, we did not include a neutral option \cite{krosnick1997rating}. In addition, participants were encouraged to volunteer information about their Employer (industry vs academia, Employer name), their role at their Employer, their highest degree and their level of seniority, as well as demographic information which included age, gender, pronouns, sexual orientation, trans identity, intersex identity, race/ethnicity and disability status. Each of these questions included an option ‘Prefer not to say’ or could be skipped. 

In total, the survey comprised a maximum of 55 questions and took approximately 11 minutes to complete. Apart from the informed consent, all questions were optional. The full survey can be found in Appendix~\ref{asec:full_survey}.

\subsection{Survey distribution} \label{met:survey_dist}
The survey was implemented on SurveyMonkey  \cite{surveymonkey} and distributed as an online form. The survey was open from November 2023 until March 2024 and specifically advertised during the Neural Information Processing Systems international conference (New Orleans, USA). The survey was heavily distributed across AI affinity groups~\cite{wiml2025, blackinai2025, queerinai, lxai2025, disabilityinai2025} 
with repeated sharing via group mailing lists as well as in-person “spin the wheel” games (workshop and social event at NeurIPS 2023). In addition to the targeting, these groups may be more sensitive to the purpose of the survey and leading them to self-select more. Therefore, we may expect higher representation of demographic minorities among survey respondents than what exists in the overall AI/ML community.

To incentivize participation of majority groups, participants could register on a separate form to win one of 10 gift cards of \$100, as well as participate in the “spin the wheel” games. The authors also personally reached out to leaders in the field (in particular at Apple, Google and Google DeepMind) to encourage their teams to participate. Based on this distribution strategy, we might expect more early career participants given the gift cards and games, participants skipping most questions to receive their prize, as well as a skew to certain Employers. We note that no Employer has officially supported this survey, e.g. by officially asking their employees to fill it out, and that most supporters were individuals in the field.

\subsection{Data analysis} \label{met:data_analysis}

\noindent\textbf{Response aggregation}: For each item, we obtain the response as a string (e.g. ‘Agree’ or ‘Never’) and compute the percentage of responses in each option. These results include “NA/I don’t know” answers and are typically represented via bar plots.

\noindent\textbf{Descriptive statistics}: To enable comparisons between subgroups, we convert our Likert scales to ordinal ranges. Responses on a 4-point Likert scale are recorded as follows: Strongly agree = 1, Agree = 2, Disagree = 3, Strongly disagree = 4. Similarly, responses on the frequency Likert scale are encoded as: Never = 1, Multiple times a year = 2, Multiple times a month = 3, Multiple times a week = 4, and Multiple times a day = 5. Responses corresponding to skipped items, “NA/I don’t know” or “Prefer not to say” were ignored in this analysis and did not contribute to the final score. Based on this encoding, the lower the score, the better the working experience for the participant.
    
Using these ordinal ranges, we estimate the median and quantiles (25\% and 75\%) of the responses to each question, which we refer to as the ‘score’. We also estimate the ‘most frequent’ response by computing the mode of responses across participants, as well as the proportion of ‘unfavorable’ responses (score $>$ 2 for 4-point Likert scale, and score $>$ 1 for frequency questions).
    
When reporting metrics across a question or category rather than on single items, we need to perform two steps of aggregation \cite{svensson2001construction}. To report the scores, we first compute the mean of the scores across items (hence obtaining one score per participant) before computing the median and quantiles across participants of this average score. For the most frequent and unfavorable metrics, we compute the metric of interest for each item and then report the median and quantiles across items. In all cases, we assume an equal weight for each item. These metrics are typically plotted as horizontal bar plots reporting the median and quantiles.

\noindent\textbf{Statistical testing}: Based on prior work, we estimate statistical significance between the scores of multiple subgroups using parametric testing. More specifically, unpaired binary comparisons refer to Student’s t-test, while paired comparisons use a paired Student’s t-test. Paired comparisons were performed when comparing changes from one question to another within participants or to compare per-question scores across a range of questions. For comparisons involving multiple categories (e.g. testing the effect of a demographic factor on a variable of interest), we perform Chi-squared tests by computing contingency tables between the two variables to study. When considering demographic factors, responses are grouped to ensure n$>$30 in each demographic subgroup, and hence improving statistical power (with a reduced degree of freedom).

\noindent\textbf{Data stratification}: These analyses are conducted across multiple splits of the data, as defined by responses to demographic questions or other characteristics (seniority, Employer). As noted previously, we exclude groups smaller than 30 (resp. 10) in the reporting across (resp. within) Employers to preserve anonymity. To increase our reporting, we combine smaller subgroups into an ‘Other’ category where feasible and further define a binarized version of each demographic attribute to represent ‘majority’ and ‘minority’ groups in our community. These group definitions are based on section \ref{sec:methods} and detailed in section \ref{sub:ow_experience}.

%% file: tables/questions_table.tex
\begin{table}[h!]
\centering
\small
\begin{tabular}{p{2.2cm}|p{5cm}}
\textbf{Category} & \textbf{Description} \\
\hline
DEI Initiatives & \textit{6 questions:} Elicit context on how an Employer approaches these initiatives (\ref{asubsec:DEI_initiatives}) \\
Belonging & \textit{5 questions:} Understand workplace engagement and culture (\ref{asubsec:belonging})  \\
Accessibility
 & \textit{4 questions:} Opt in for those who identify as individuals with a disability or other chronic condition to understand availability of accommodations (\ref{asubsec:accessibility}) \\

Microaggressions &
\textit{15 questions:} Opt in, to understand subtle, cumulative and unpleasant situations (\ref{asubsec:microaggressions}) \\ 

Misconduct
& \textit{4 questions:} Understand an Employer's reaction to situations that go wrong (\ref{asubsec:misconduct}) \\

Performance and Compensation
 & \textit{7 questions:} Understand (dis)satisfaction with how Employers evaluate performance and determine compensation (\ref{asubsec:performance_compensation}) \\

Growth
  & \textit{4 questions:} Opportunities provided by the Employer (\ref{asubsec:growth}) \\

Well Being
 & \textit{5 questions:} Understand capacity and support for work-life balance (\ref{asubsec:well_being}) \\

Overall Satisfaction
& \textit{5 questions} (1 frequency question): Understand workplace (dis)satisfaction at a high level (\ref{asubsec:overall_satisfaction}) \\

\end{tabular}
\caption{Survey question description}
\label{table:survey_questions}
\end{table}

%% file: sections/04_results.tex
\section{Results}\label{sec:results}

A total of 1260 participants filled out the survey, with an average completion time of 11 minutes and a completion rate of 73\%, leading to 796 complete responses. Participants were split between academic and industry employers, roles, and seniority levels (table \ref{table:survey-demographics}). 

\input{tables/table1}

\input{tables/table2}

\begin{figure*}[!htbp]
    \centering
    \includegraphics[width=0.8\textwidth]{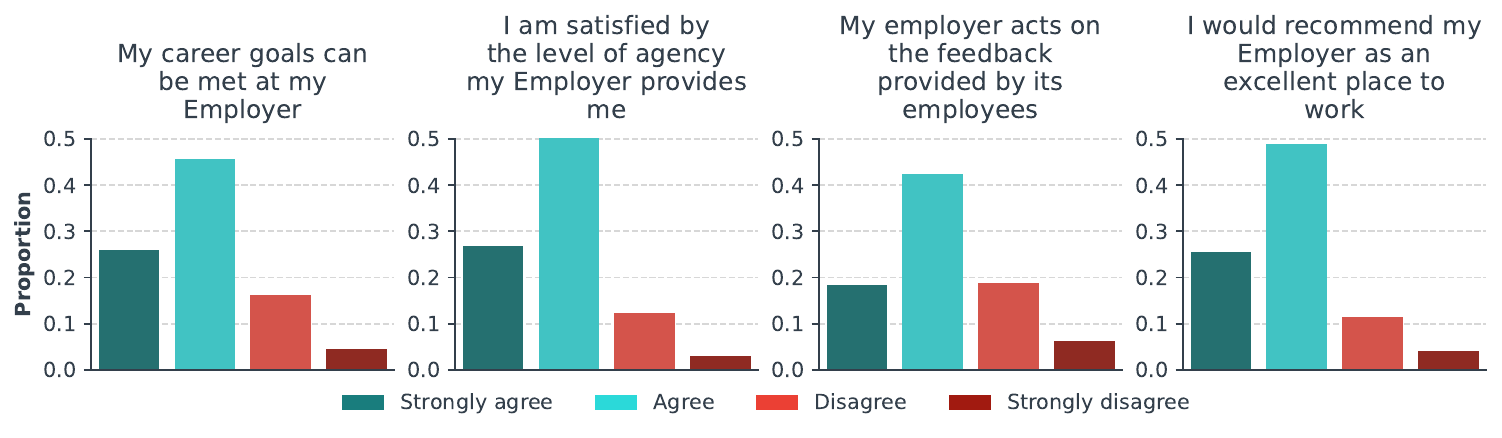}
    \caption{Responses to “Overall satisfaction” items: favorable responses are color-coded with different shades of teal, while unfavorable responses are color-coded in shades of red.}
    \label{fig1a:overal_satisfaction}
\end{figure*}

Employers were located in North America (n=369, with USA: n=335, Canada: n=34), Europe (n=180, with UK: n= 71), Asia (n=70), Africa (n=46), South America (n=8), Australia (n=5) and unknown locations (undisclosed: n=20, no answer: n=98).

\subsection{Representation of demographic subgroups} \label{subsec:rep-demo}
Table \ref{table:demographics} displays the representation of participants across the different demographic questions asked at the end of the study, for which there are at least two groups with n$>$30. We note that this limit prevents analyzing the data across participants who reported being Trans (n=25, non-disclosure=40) or Intersex (n=6, non-disclosure=40). Where necessary, we aggregate multiple subgroups within a demographic dimension to reach a minimum of n=30. These combined groups include: 46+ (46-55 (n=35), 56-65, 65+), other genders (non-binary, non-binary woman, gender non-conforming, genderfluid, genderqueer, agender, questioning), other pronouns (they/them, any/all, she/they, he/they, xe/xem, questioning), other sexual orientations (gay, lesbian, pansexual, asexual, queer, demisexual, pansexual and ace, questioning), other races and ethnicities (South East Asian, Native American/Alaska Native/First Nations, Pacific Islander, other). When we further filter for other factors or consider intersectionality, we binarize each identity dimension into its majority and minority subgroups.

While the representation among most demographic factors matches our expectation and previous reports in our community, we note that women are over-represented in this sample (60.55\% compared to around 22\% in the field~\cite{InterfaceAI}). This over-representation is also present for pronouns (60.30\%). Therefore, this can be a confounding variable when considering other demographic identities. In fact, most of our respondents are women who identify as White, East Asian, or South Asian, with twice as many women respondents in these races/ethnicities compared to men. Therefore, where possible, we split the results by gender when discussing race. On the other hand, there are similar proportions of able-bodied and disabled men and women.

\subsection{Overall workplace experience} \label{sub:ow_experience}

Here, we describe the overall responses obtained for each item, highlighting an overall favorable response to our questions, while presenting variability across respondents and items. We first analyze the responses to the “Overall satisfaction” items in figure \ref{fig1a:overal_satisfaction}.

\begin{figure}[!htbp]
\centering
\includegraphics[width=0.9\columnwidth]{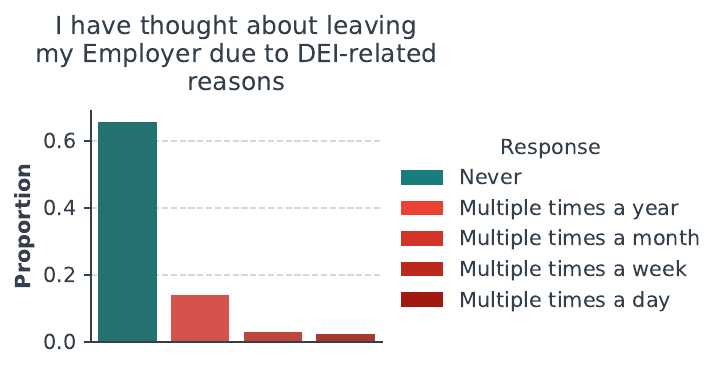} 
\caption{Proportion of the frequency with which employees consider leaving their employer due to DEI-related reasons. Favorable results are depicted in green while unfavorable results are displayed in shades of red.}
\label{fig1a:freq_leaving}
\end{figure}

\begin{figure*}[!htbp]
\centering
\includegraphics[width=\textwidth]{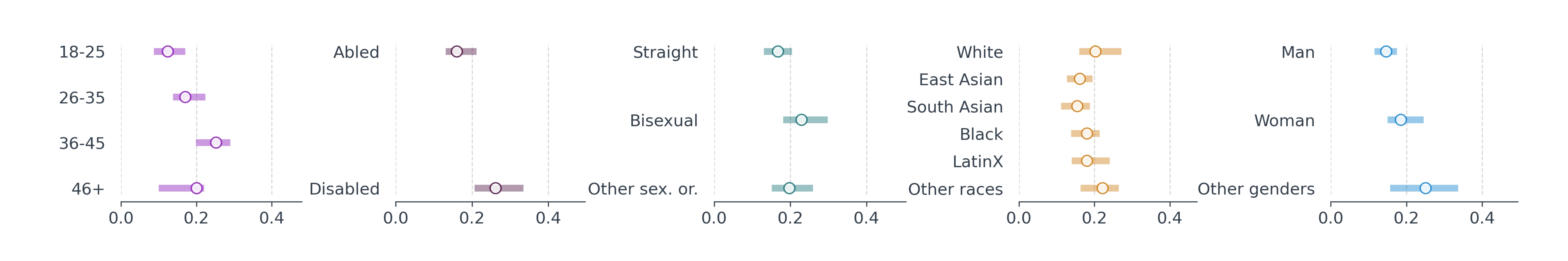} 
\caption{Fraction of unfavorable responses by “age”, “disability”, “minority sexual orientation”, "gender" and "race".}
\label{fig:demographic_unfavorable}
\end{figure*}

In figure \ref{fig1a:freq_leaving} we analysed the last item of this category which asks for a frequency response to: “I have thought about leaving my Employer due to DEI-related reasons”, with options including ‘Never’, ‘Multiple times a year’, ‘Multiple times a month’, ‘Multiple times a week’ or ‘Multiple times a day’. While 65.70\% of respondents never considered leaving their Employer for DEI reasons, 19.10\% of respondents considered leaving for DEI reasons at least multiple times a year (year: n=111, month: n=23, week: n=18, day: n=0). Among those 152 Employees, there is a similar proportion in academia (53.63\%) and in industry (42.34\%), and the career status of the respondents matches that of our participant pool (working towards degree=35.53\%, early career=32.24\%, mid-career=25\%, senior=7.24\%). When considering demographic factors, a large proportion of respondents who considered leaving their Employer for DEI reasons at least multiple times a year mostly identified as women or other minority genders (76.97\%, compared to 69.97\% of overall respondents, chi-squared test on effect of gender groups on responses: $p$=0.0579, df=6). We note an increase in the proportion of respondents identifying with a minority sexual orientation in this sample compared to our participant pool (26.32\% compared to 20.60\% in total, chi-squared test: $p$=0.2556, df=6), as well as a small increase in respondents identifying as Trans or Intersex. Compared to our sample of respondents, there was no major deviation in proportion in terms of race or ethnicity. With regards to disability status, participants who reported a disability had significantly higher responses to this question than those who did not report a disability (Student’s t test: $p$=0.000048).

Overall, we observe that the respondents were satisfied with their Employer in terms of career goals and agency, and would recommend their Employer as an excellent place to work. However, we note a less favorable response when considering how the Employer addresses feedback provided by its employees. Lastly, when we estimate aggregate metrics across 4-point Likert scale items in DEI Initiatives (5 items), Belonging (5 items), Misconduct (4 items), Performance and Compensation (7 items), Growth (4 items), Well Being (5 items) and Overall Satisfaction (4 items) questions. We note that each item obtained a most frequent response (mode) and a score (median) of 1 or 2, indicating overall satisfaction with the workplace experience. However, each item also displayed responses covering the full range of options, from strongly agree (=1) to strongly disagree (=4).

Based on these results, it is likely that different demographic subgroups do not share similar workplace experiences, which aligns with past research in different fields \cite{plaut2011perceptions, mckay2008mean, ely2001cultural}.

\subsection{Workplace experience across demographics} \label{sub:workplace-exp}
In this section, we report different metrics across demographic subgroups, aggregating in minority vs majority groups where necessary. We consider the same 4-point Likert scale items as in the previous section and present the results in terms of the fraction of unfavorable responses given its higher granularity.

As displayed in figure \ref{fig:demographic_unfavorable}, we observe that respondents identifying with different genders have different rates of unfavorable responses, with men being overall more favorable than women, and other genders reporting more unfavorable responses. We also note a larger variability of this fraction of unfavorable responses across items for women and non-binary genders. Similarly, we observe that the fraction of unfavorable responses increases with age, having a disability or chronic condition, or a minority sexual orientation, as shown in the figure \ref{fig:demographic_unfavorable}. Due to the more negative experiences of women and their over-representation in this sample, we observe similar rates of unfavorable responses for White respondents as for some racial or ethnic minority groups. White respondents aside, we notice that East and South Asian participants display lower fractions of unfavorable responses than Black, LatinX, and Other races, as shown in figure \ref{fig:demographic_unfavorable}.

\begin{figure}[!htbp]
\centering
\includegraphics[width=\columnwidth]{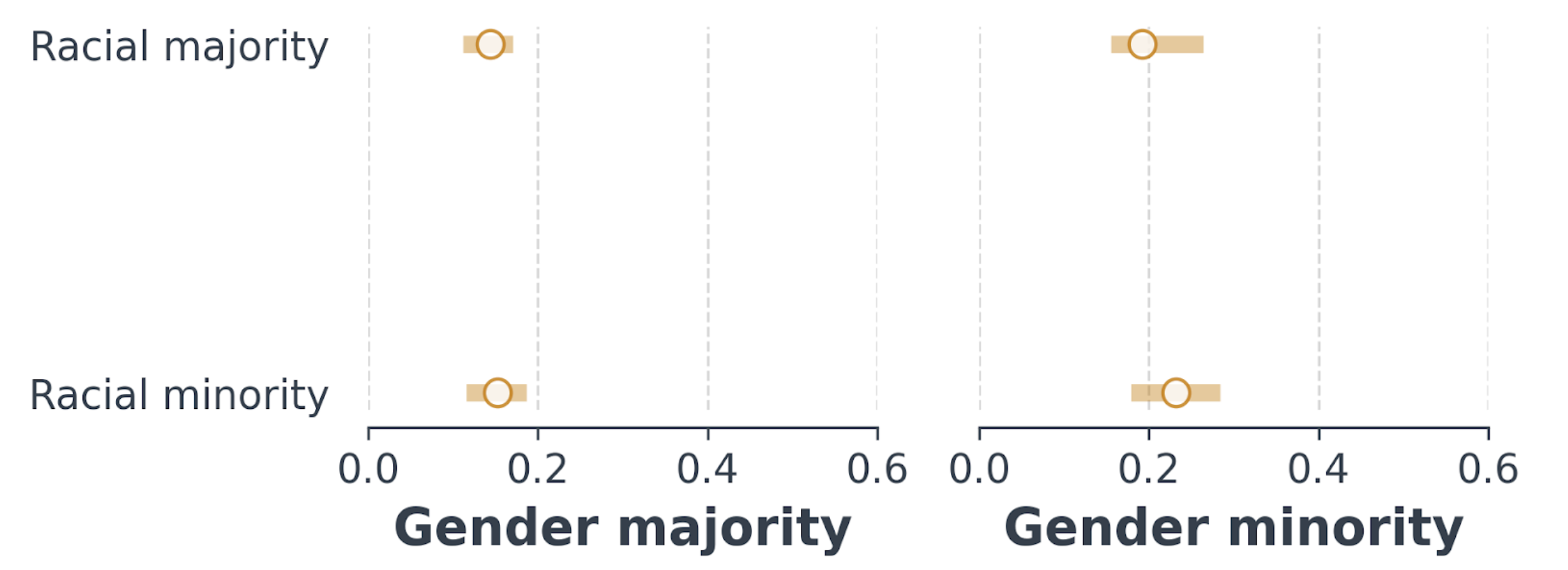} 
\caption{Rates of unfavourable answers by racial majority and minority across gender majority and minority.}
\label{fig1:min_maj_min}
\end{figure}

We hence split per gender, but have to binarize the categories to ensure n$\geq$30. We then notice the overall trend that men and gender minorities report different rates of unfavorable answers, and that there is an effect of race and ethnicity as well, especially for the gender minority, as shown in figure \ref{fig1:min_maj_min}. This result is aligned with multiple prior research highlighting how white women and women of colour do not share the same experiences \cite{crenshaw1991mapping, settles2006intersectional, wingfield2010emotions}.

\begin{figure*}[!htbp]
\centering
\includegraphics[width=0.8\textwidth]{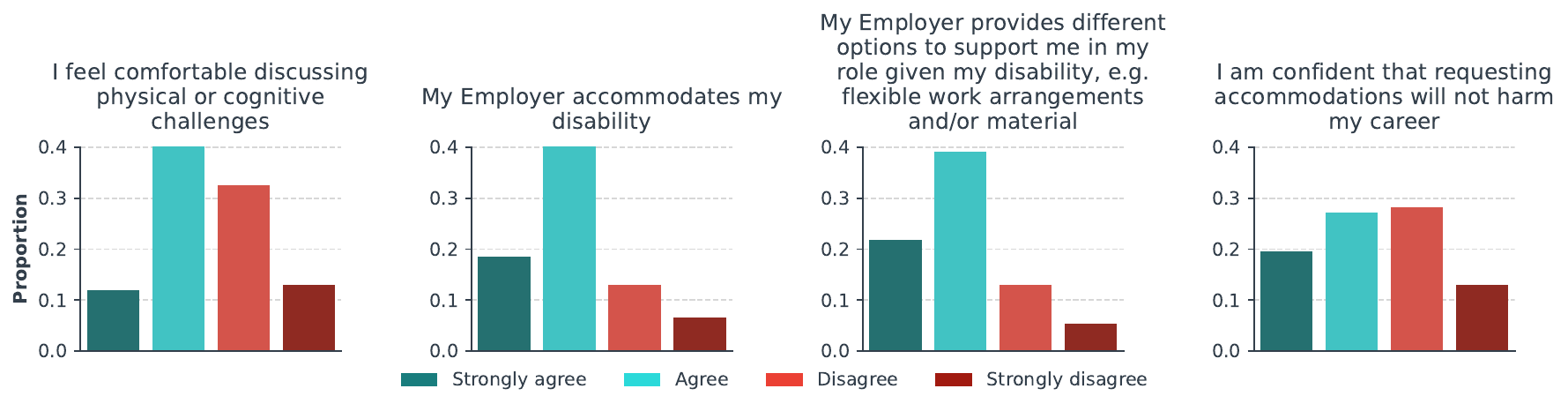} 
\caption{Proportions of each response to Likert-scale items related to accessibility and disability. Favorable answers are displayed in a green gradient, while unfavorable answers are displayed in a red gradient..}
\label{fig1:accessibility}
\end{figure*}

\subsection{In-depth analysis}

In this subsection we provide further in-depth analysis on some specific question categories and minority groups.

\subsubsection{Accessibility and disability}
Participants who identified as disabled or with a chronic condition (n=92) were presented with further survey items relating to the accessibility and the management of their disability in the workplace. According to figure \ref{fig1:accessibility}, we note that a majority of respondents did not feel comfortable discussing physical or cognitive challenges with their Employer (45.65\%, n=92), and that an important proportion are not confident that requesting accommodations will not harm their career (41.30\%, n=81).

\subsubsection{Microaggressions
}
We first asked a general question about microaggressions: “I experience microaggressions because I am part of an underrepresented group.” Among the 796 total respondents, 327 declared that they never experienced microaggressions, while 384 experienced microaggressions in the workplace. Amongst those experiencing microaggressions, 81.51\% identified as part of a gender minority (n=313, Chi squared test: $p$ $<$ 0.00001, df=4), 30.99\% identified with a minority sexual orientation (n=119, Chi squared test: $p$ $<$ 0.00001, df=4) and 15.63\% have a disability or chronic condition (n=60, Student’s t test: $p$ = 0.0000117). These proportions are higher than those of the pool of respondents. We noted a slight difference in terms of age (n=99 of respondents 35+, i.e., 25.78\% compared to 23.49\%) which is not significant (Chi squared test: $p$=0.8057, df=4).

\begin{figure}[!htbp]
\centering
\includegraphics[width=\columnwidth]{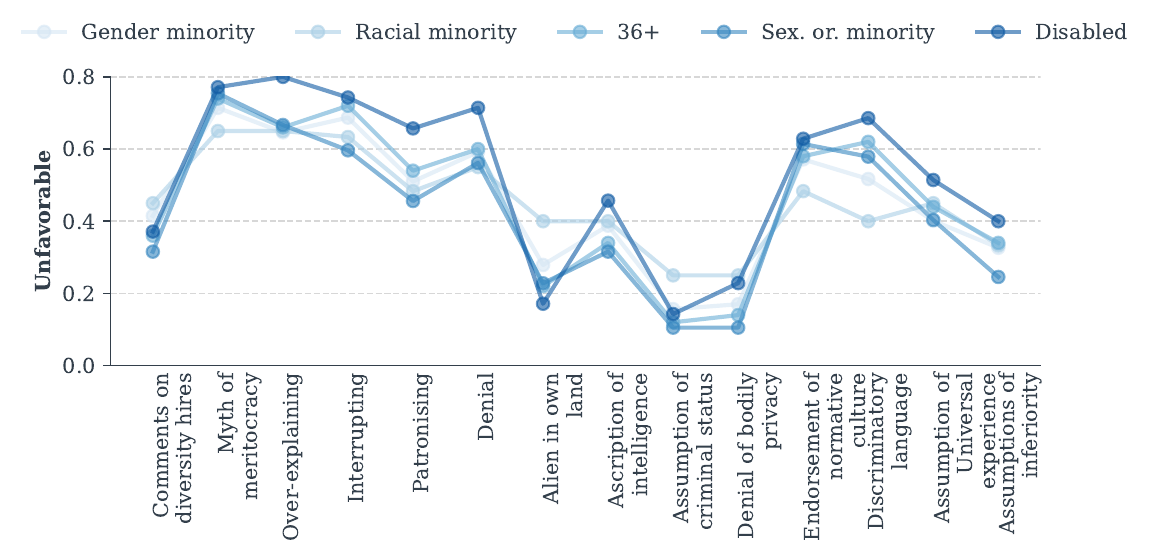} 
\caption{Rates of unfavorable responses in each demographic minority (please note potential overlap of respondents due to intersectionality) across the detailed items related to microaggressions. Each demographic minority is represented by a color.}
\label{fig1:graph}
\end{figure}

After providing consent, some respondents (n=176) provided further details on the types of microaggressions they experienced, and how frequently. Based on the logic of our survey, we can expect that most respondents have experienced microaggressions, which is confirmed by the low number of respondents from majority groups (e.g, only 23 men), and the overall unfavorable score of 45.74\% across these items. We hence focus on which microaggressions were experienced most frequently across demographic groups. These are displayed in figure \ref{fig1:graph}, in which we observe a strong correlation between the proportions of unfavorable responses across minority groups. As exposed in prior sections, we observe the highest rates of unfavorable responses for respondents identifying as disabled or with a chronic condition. On the other hand, we observe higher unfavorable rates for the racial minority when it comes to microaggressions related to assumptions of being an alien in their own land, with assumptions of criminal status, and with a denial of bodily privacy. These have been documented previously, and can take the form of consumer racial profiling~\cite{gabbidon2007consumer,feagin2018continuing, schreer2009shopping} checking badges more often for employees of racial or ethnic minorities, or questions about hairstyles~\cite{trusty2023hair, summers2022hair}. Finally, we also observe a high rate of experiences related to the endorsement of normative culture and behaviors reported by members of sexual minorities~\cite{corlett2023manifestations, woodruffe2016countering, bizzeth2023ah}. These may translate as assuming someone’s partner is from the opposite sex, or depicting families as a heterosexual couple.

    \begin{figure*}[htbp]
    \centering
    \includegraphics[width=\textwidth]{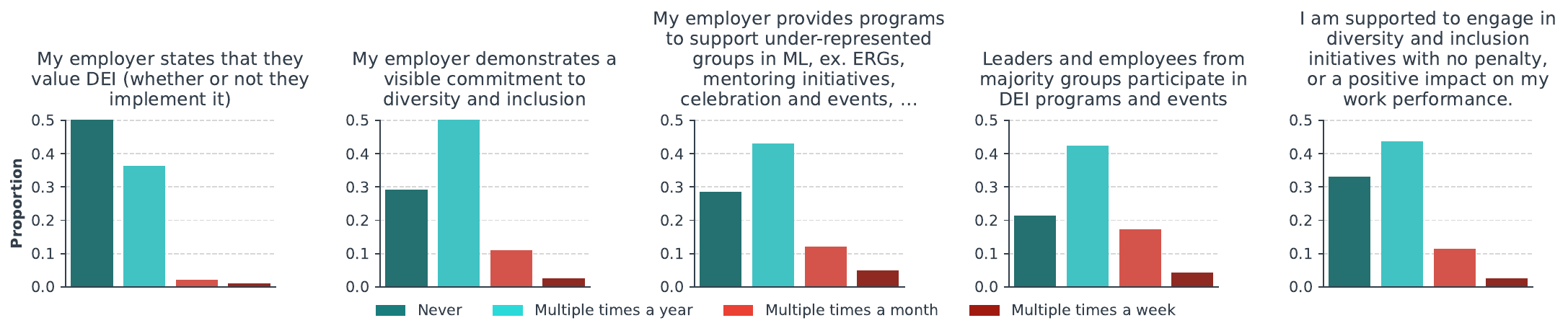} 
    \caption{Proportions of responses across DEI Likert items.}
    \label{fig1:dei_items}
    \end{figure*}

A surprising result indicates that 28.91\% (n=111) of respondents experiencing microaggressions identified as part of a racial or ethnic minority, which is similar to the ~30\% of overall respondents. However, as noted in previous sections, this result is biased by the over-representation of racial majority women (especially of White women) and may hide the experiences of specific subgroups.

\subsubsection{Walking the Walk or just Talking the Talk?}
Next, we probe different aspects of each Employer’s commitment to DEI as shown in figure \ref{fig1:dei_items}. For instance, the first item asks “My employer states that they value DEI (whether or not they implement it)” while the second item asks “My employer demonstrates a visible commitment to diversity and inclusion”. We observe a significant increase in unfavorable responses across these questions (paired Student’s t test: $p$ $<$ 0.00001).

84 participants reported that their Employer does not have DEI programs. We also find that DEI initiatives are mostly led by members of minority groups (n=533), and that leaders and members of majority groups could participate more in these events and initiatives (n=304).

\subsection{Employers}

The participants reported a variety of Employers, both in academia (n=235) and in industry (n=140). While we post-processed the data to consolidate responses where the ‘Other’ response was selected that shared the same Employer, most Employers are represented by only a few respondants to our survey, reflecting challenges. In academia, the Massachusetts Institute of Technology had the most participants (n=12), followed by Carnegie Mellon University (n=11) and University of Pennsylvania (n=8). Based on those low numbers of respondents, we do not conduct further analysis for academic Employers.

\begin{figure}[!htbp]
\centering
\includegraphics[width=0.7\columnwidth]{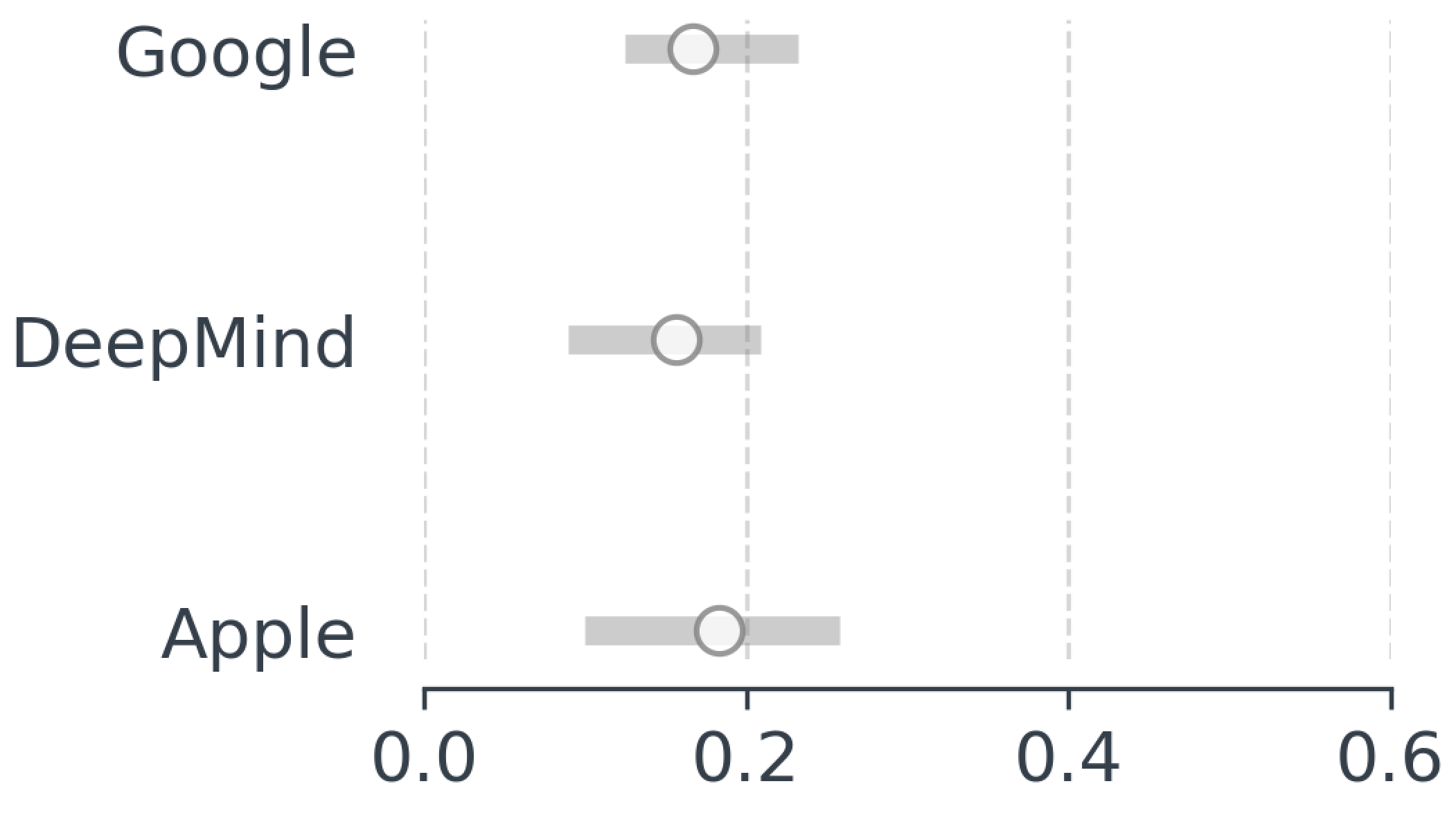} 
\caption{Overall fraction of unfavourable responses by employer.}
\label{fig:overall_employer}
\end{figure}

In industry, a small number of Employers have participation rates above our anonymization cutoffs, leading to the computation of unfavorable scores for Google DeepMind (n=67), Google (n=42) and Apple (n=30) as shown in figure \ref{fig:overall_employer}. We note that Google DeepMind has slightly better scores than other companies, but a paired t-test on the proportion of unfavorable responses per question displays that this difference is not statistically significant ($p$=0.0823) between DeepMind participants and other Employers. Comparing Google (resp. Apple) to the rest of Employers leads to similar results ($p$=0.7668, $p$=0.7547 respectively).

\begin{figure}[!htbp]
\centering
\includegraphics[width=\columnwidth]{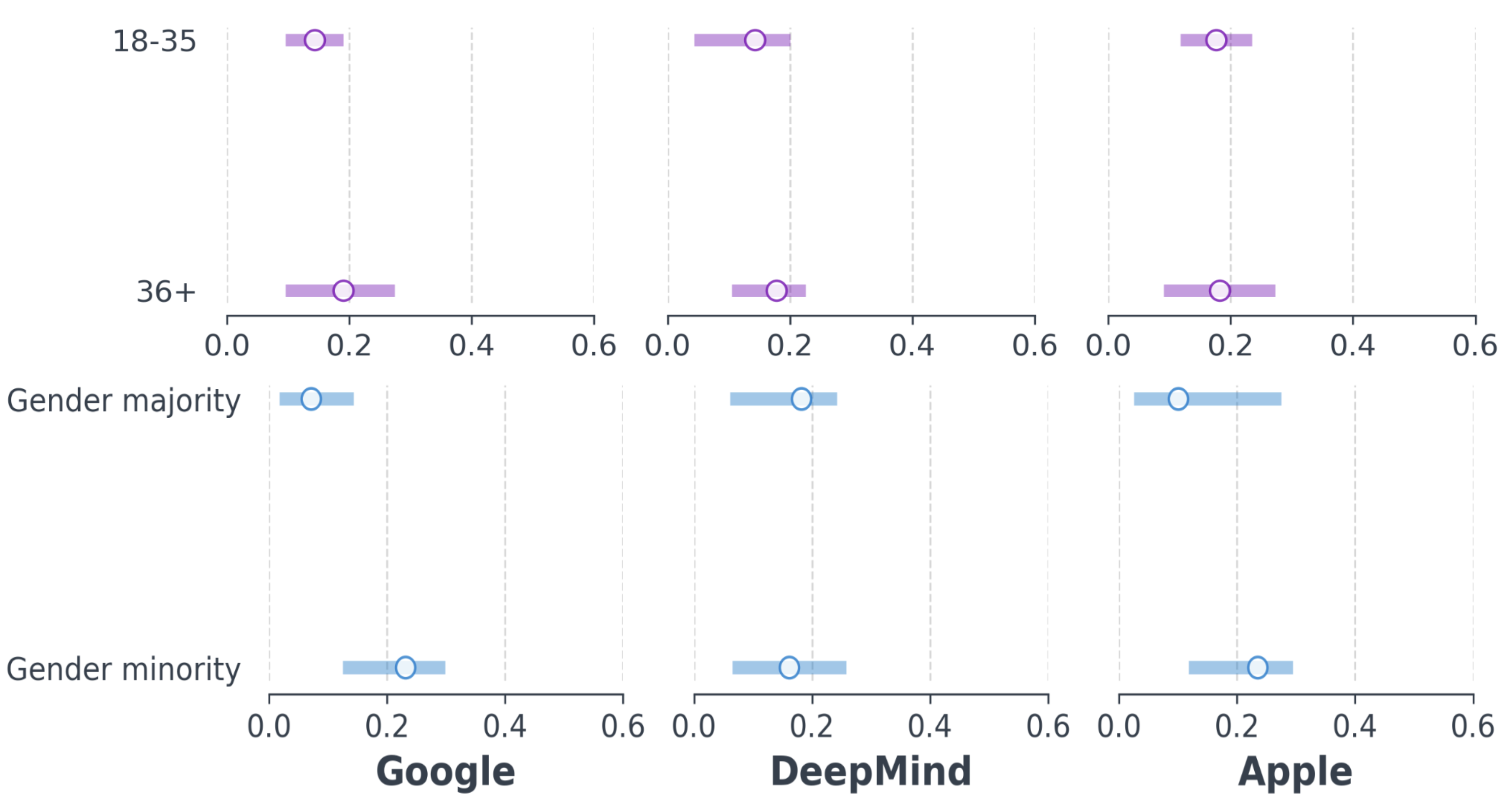} 
\caption{Fraction of unfavourable responses by employer per gender (blue), and per age.}
\label{fig:gender_age_employer}
\end{figure}

When investigating the experiences of specific demographic subgroups at these Employers, we can only report on age and gender given our minimum n=10 (as per IRB). We note different patterns across Employers (figure \ref{fig:gender_age_employer}), but prefer not to draw conclusions given the low numbers of participants.

\begin{figure*}[!htbp]
\centering
\includegraphics[width=0.95\textwidth]{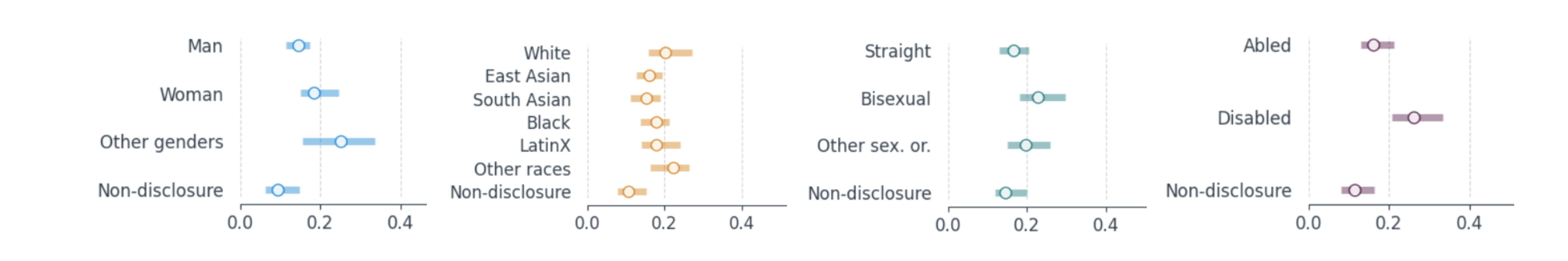} 
\caption{Unfavorable scores across respondents with undisclosed identities to those of respondents with disclosed identities.}
\label{fig1:non_disclosure}
\end{figure*}

\subsection{Non-disclosure}
In this subsection we describe the ``non-disclosure" across different aspect of the survey. Due to the sensitivity that can be perceived from the different categories, we believe this aspect is worth reporting and being represented as part of the findings.
\subsubsection{Non-disclosure of Employer}
Across the 796 participants who completed the survey, only 39 ($\sim$5\%) preferred not to disclose their Employer, across academia (n=16) and industry (n=23). We note that some industry participants belonging to small organizations referred to general terms as ‘startup’ instead of naming their Employer, and they are included as non-disclosure. In academia and industry, non-disclosure of Employers was related to younger participants (26-35: n=21, 36-45: n=9, Prefer not to say: n=3, 18-25: n=2, 46-55: n=1, unanswered: n=3) who identified as women (women: n=22, men: n=10, Prefer not to say: n=2, other genders: n=2, unanswered: n=3). There were no trans or intersex participants in this sample, and sexual orientation did not reveal a specific trend (straight: n=20, Prefer not to say: n=9, bisexual: n=3, other sexual orientations: n=2, unanswered: n=3). In terms of racial or ethnic identity, the majority of non-disclosures came from participants identifying as White (n=13) or South Asian (n=10). These low numbers display an overall trust in the anonymity of our survey, although Employees of small startups would benefit from further reassurance that the data is not shared with their Employer.

\subsubsection{Non-disclosure of demographics}
The number of ‘Prefer not to say’ responses per demographic identity item is reported in table \ref{table:demographics}. We note that participants who did not disclose an identity were less likely to disclose any of them, with the highest proportion of non-disclosure for the item asking about sexual orientation. When considering undisclosed sexual orientation, we note higher rates of undisclosed gender (n=24, compared to n=32 in the participant pool of 796 respondents), age (n=14, compared to n=18), trans (n=27, compared to n=38), or intersex identity (n=28, compared to n=40). In terms of race and ethnicity, we also note a higher rate of undisclosed race (n=39, compared to n=65 in total). However, the racial and ethnic distribution of respondents with undisclosed sexual orientation but disclosed race/ethnicity is markedly different from that of the overall population, with Black and African-American respondents (n=17) being represented similarly to White respondents (n=18). Given differing legal contexts in each country when it comes to sexual orientation, we also look at the employers’ geographical location of respondents with undisclosed sexual orientation. We observe the following distribution: USA (n=34), Europe (n=13), Asia (n=11), Africa (n=10), South America (n=2), undisclosed (n=11), no answer (n=29). This result shows an over-representation of respondents with Employers in Africa who prefer not to disclose their sexual orientation. We, however, note that none of these subgroups of undisclosed sexual orientation and minority race/ethnicity or geographical location include 30 respondents.

When n$\geq$30, we review the scores across Likert items of respondents with undisclosed identities to those of respondents with disclosed identities. For gender, race, sexual orientation, and disability status, we observe that the overall scores of respondents with non-disclosed identities are lower than any other identity, including the majority identity, as shown in figure \ref{fig1:non_disclosure}. This latter analysis suggests that non-disclosure is mostly related to an overall non-disclosure of identities and to respondents skipping items, which might seem to align with more favorable workplace experiences and could reflect a majority group (i.e., respondents who do not identify with any marginalized or underrepresented identities).

These results display an overall trust in the anonymity of our survey, and that respondents with marginalized identities were comfortable sharing their experiences in the workplace. We, however, note that respondents were less likely to share their sexual orientation, especially if they identified as Black or African-American. When comparing scores, we note that respondents with undisclosed identities had lower scores than any identity (including majority groups in each identity), and we suspect that these respondents felt less concerned by the purpose of the survey and might have been drawn by the incentives to complete the survey (prize draw or spin the wheel) rather than by a desire to share their workplace experience.

%% file: tables/table1.tex

\begin{table}[h!]
\centering
\small
\begin{tabular}{p{1.2cm}|p{4cm}|p{2.5cm}}
\textbf{Question} & \textbf{Option} & \textbf{Proportion (n)} \\
\hline
Employer & Academia & 55.53\% (n=442) \\
 & Industry & 44.47\% (n=354) \\
\hline
Seniority & Working towards degree & 30.53\% (n=243) \\
 & Early career (0--4 years) & 33.04\% (n=263) \\
 & Mid-career (5--15 years) & 29.15\% (n=232) \\
 & Senior (15+ years) & 6.66\% (n=53) \\
 & Undisclosed & 0.62\% (n=5) \\
\end{tabular}
\caption{Survey respondent demographics}
\label{table:survey-demographics}
\end{table}

%% file: tables/table2.tex
\begin{table*}[h!]
\centering
\small
\begin{tabular}{l|l|l|l|l|l|l}
 & \textbf{Age} & \textbf{Gender} & \textbf{Pronouns} & \textbf{Sexual Orientation} & \textbf{Race/Ethnicity} & \textbf{Disability} \\
\hline
\textbf{Majority} & 26-35 (n=441) & Man (n=239) & He/him (n=230) & Straight (n=494) & W (n=314) & No (n=638) \\
 & 18-25 (n=141) & & & & EA (n=118) & \\
 & & & & & SA (n=117) & \\
\hline
\textbf{Minority} & 36-45 (n=137) & Woman (n=482) & She/her (n=480) & Bisexual (n=83) & AAB (n=89) & Yes (n=92) \\
 & 46+ (n=50) & Others (n=32) & Others (n=44) & Others (n=81) & H/L (n=50) & \\
 & & & & & ME (n=52) & \\
 & & & & & OR (n=54) & \\
\hline
\textbf{Prefer not to say} & n=18 & n=32 & n=31 & n=110 & n=65 & n=61 \\
\end{tabular}
\caption{Demographic representation of survey participants across multiple dimensions. Each subgroup presented includes at least 30 participants. W - White, EA - East Asian, SA - South Asian, AAB - African American or Black, H/L - Hispanic/LatinX, ME - Middle Eastern, OR - Other Races.}
\label{table:demographics}
\end{table*}

%% file: sections/05_discussion_conclusion.tex
\section{Discussion}\label{sec:discussion}
The AI Community Pulse project was initiated in 2022, with the survey running between November 2023 and March 2024. The survey focused on understanding overall workplace experiences, by asking about an individual's experiences related to the following constructs: Belonging, Misconduct, Performance and compensation, Growth, Well-being, Overall satisfaction and Microaggressions (see Table~\ref{table:survey_questions}). The results indicate that there are disparities in workplace experiences for underrepresented and marginalized demographics. In particular, we highlight that accessibility remains an important challenge for a positive workplace experience, and that disabled employees have a worse experience in the workplace than their non-disabled colleagues. We discuss our findings more generally in the subsequent sections. 

\textbf{DEI data in industry and academia is notoriously hard to find.} Industry diversity data is incomplete and fragmented, and often only reported after pressure from civil society ~\cite{bostonglobe_google_audit, washingtonpost_google_audit}. In the US, industry is currently abandoning DEI efforts, making current data even more difficult to find. Our survey begins to fill these growing gaps, by engaging AI/ML professionals from multiple institutions (both in academia and industry) as from multiple demographics. However, despite significant efforts, this data was still challenging to collect, requiring coordination with multiple AI affinity groups.

\textbf{DEI problems have not been addressed despite many efforts.} Our studies show persistent gaps in workplace experiences, finding disabled people, women and gender minorities, women of color, and people who aren't straight have worse experiences with employers, and experience higher rates of microaggressions. Our study shows that past DEI efforts have not been successful at eliminating the gap. This is especially concerning given the future of DEI programs in light of the recent backlash against DEI measures in the US and in the ensuing reduction of DEI measures in companies and institutions~\cite{Sherman2025backtrackdei, Cai2025Googlebacktrackdei, MurrayBohannon2025IBMbacktrackdei}.

\textbf{Bias and discrimination in the workplace mirror bias and discrimination in AI.} It is well-known that current state-of-the-art models are biased, with minority groups disproportionately affected by the deployment of these models in downstream applications~\cite{davis2021algorithmic}. For example, representational harms~\cite{weidinger2021ethical, weidinger2023sociotechnical}, are harms that arise from AI outputs and reinforce negative stereotypes as well as discrimination. When these harms intersect with occupational use cases of AI~\cite{hall2023visogender, kirk2021bias}, the potential to perpetuate discrimination and a lack of diversity in the workplace is high. With these factors affecting teams, and the development of responsible AI~\cite{papagiannidis2025responsible}, there becomes a self-perpetuating cycle of bias and harm in AI development. Further, considering these models all dictate access to resources (allocational harms~\cite{weidinger2021ethical, weidinger2023sociotechnical} and quality of service harms~\cite{shelby2023sociotechnical}), it is important to critically assess the fact that AI is deployed in student admissions processes where there is evidence of bias in these outcomes~\cite{van2023analysis}. These biased and discriminatory outcomes could further limit representation in STEM higher education~\cite{walton2019belonging, xu2015barriers}. In turn, this will further narrow the AI/ML professional talent pool, creating the same, aforementioned self-perpetuating cycle. 

\textbf{We need to develop good ways to understanding experiences of AI workers.} This study is of importance as it provides a foundation for tracking the effectiveness of DEI initiatives over time. This is especially important given that the aforementioned significant socio-political shift, and a significant backlash against DEI efforts~\cite{WhiteHouse2025}, particularly within the US context which is where a significant proportion of the participants are based (46\%). The experiences of individuals in the workplace are likely to change along with this shift, and this is likely to lead to more unfavorable outcomes given that our study found that Employees who identified with minority groups were key in pushing DEI initiatives in the workplace. Given the push back on DEI and affirmative actions, we believe it is important for all Employers to be transparent and support initiatives that provide a window into the work experiences of their employees and on how they compare to other organizations in these dimensions. The very same demographics experiencing disproportionately unfavorable workplace experiences, are also disproportionately negatively impacted by discriminatory biases in AI models. The teams and the models they are developing are clearly suffering from the lack of improved representation. Not only could the commitments to reducing disparities in the workplace reflect the AI/ML commitment to equity and justice, but studies have shown that improvements in DEI have resulted in higher standards of responsible development~\cite{papagiannidis2025responsible} and therefore, better AI models for all. Therefore, understanding these disparities in workplace experiences is paramount.

\subsection{Limitations}
While an important step forward in surfacing disparities in workplace experience, the study is not without limitations.  Firstly, we were unable to perform some comparisons safely as our survey is not an unbiased sample of AI as a whole. Therefore, it does not tell us about rates at which various groups participate in AI work. Notably, a significant proportion of responses came from women, i.e. 60.55\% compared to around 22\% in the field~\cite{InterfaceAI}. On the other hand, members of majority groups such as white men did not engage with some aspects of the survey (e.g. men in detailed microaggressions), preventing comparisons with minority groups. Motivating participation in surveys is an ongoing, broad challenge~\cite{Porter2004surveyfatigue, Artino2022surveyfatigue}. Further, the participants are primarily from Western perspectives, with 46\% from North America (primarily the US) and the next highest proportion of participants (23\%) coming from Europe (UK, primarily). Therefore our analysis is primarily in a Western (US, Canada, UK, Europe) context and we assess the majority as ``most favored'' groups in this context, compared to minority groups. This context dictated our definitions of majority and minority groups or other combinations of demographic identities and requires justification of our use of binarization to support statistical analysis. We recognize this can obfuscate valid granularity in experience. 

Over-participation can be related to the distribution of the survey in affinity group mailing lists and events, which attract more members from underrepresented groups. Under-participation can be attributed to the dissemination techniques, the lack of representation within these spaces, and a lack of buy-in from Employers themselves. In our experience, only a few Employer leaders expressed their support, and even then only through personal, not employer, channels. While one of the primary goals was to provide a platform for prospective Employees to assess how they might experience their workplace at different Employers, we have not been able to garner enough support of academic institutions or companies to do so safely.

\subsection{Future work}

Future dissemination of the survey will require dedicated efforts to engage more majority groups, for instance, partnering with companies and universities, as well as having active efforts at multiple conferences, on multiple continents to engage more participants from different demographics. Given the rapid change in socio-political climate during the lifespan of this survey, special efforts would be needed to surface the experience of certain groups, such as individuals who identify as transgender~\cite{Thoreson2025TrumpAdmin}, who may experience increased fear in disclosing their identities in the workplace~\cite{paine2025toomuchtrouble}, as well immigrants and non-citizens. Future studies may also include concerted efforts to increase the representation of participants from more countries as the literature demonstrates that disparities in workplace experiences exist in other regions globally~\cite{maji2024they, de2024present}. Future surveys may surface granularity in perspectives through qualitative methods. Our survey demonstrates that their is an interest in the community to engage, as we report through analysis of the open text responses (see Appendix~\ref{asec: additional_findings} for this analysis). Finally, in the wake of the socio-political changes around DEI, we hope to repeat versions of this study to create longitudinal data to enable us to study the effectiveness of different DEI interventions.

%% file: sections/98_acknowledgements.tex
\section*{Acknowledgements}
This research could not have been done without the contributions of many people. We would like to acknowledge the following individuals for their help and support in the design, data collection and writing process: Maria Skoularidou, Tolani Britton, Rediet Abebe, Gelyn Watkins, Sanmi Koyejo, Wendy DuBow, Alana Anderson, Lecia Barker, Robin Brewer, Joanne Esch, Nina De Hora, Naniette Coleman, Bethany Edmunds, Alessandra Tosi, Caroline Weis, Merve Gurel, Erin Grant, Laura Montoya, and Amal Rannen-Triki. This work was funded by WiML, with support from Black in AI.

%% file: sections/99_appendix.tex
\section{Additional Findings}\label{asec: additional_findings}

\subsection{Workplace experience} 
\begin{figure*}[h!]
\centering
\includegraphics[width=0.8\linewidth]{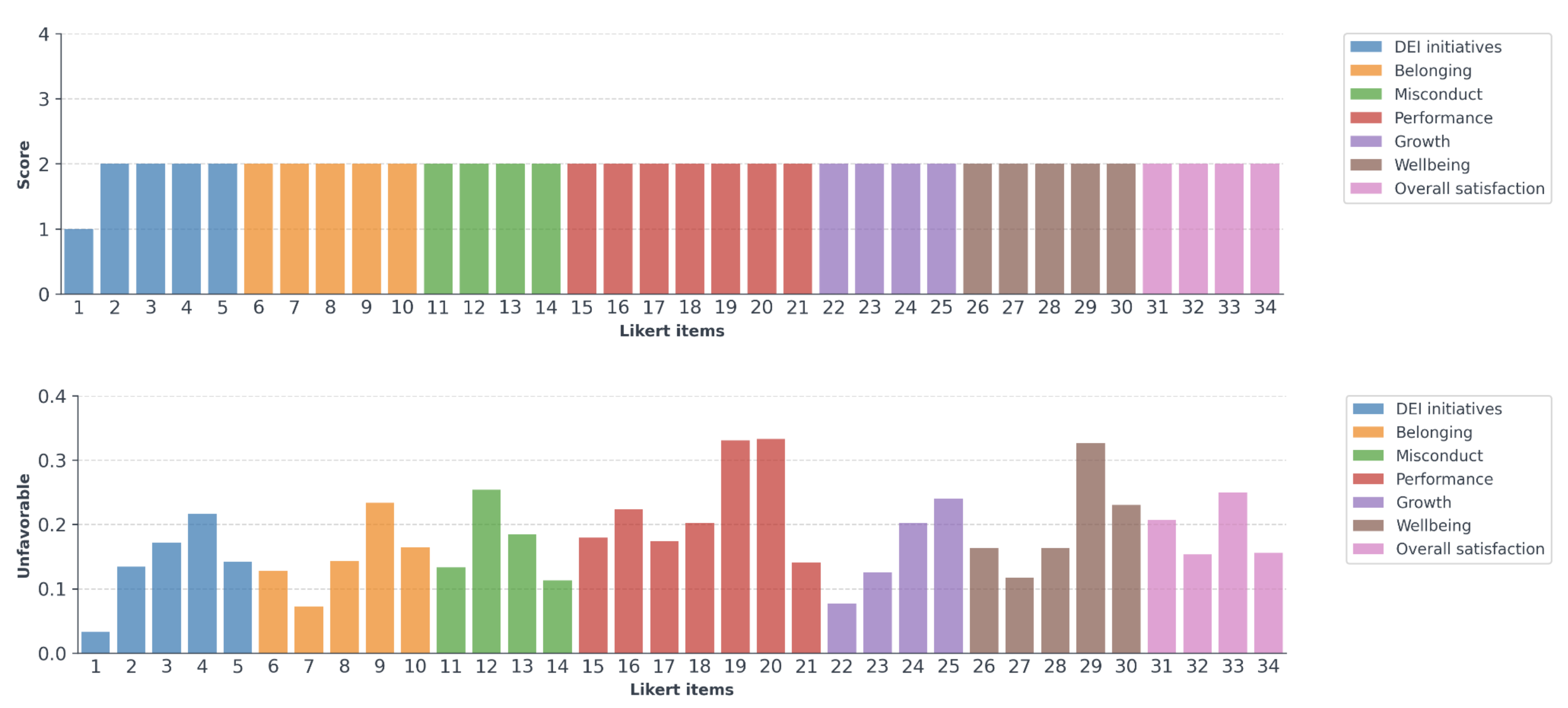} 
\caption{(top) Median response (bottom) Fraction of unfavourable responses for each Likert item, color-coded by category.}
\label{fig1:median_unfav}
\end{figure*}

For each item, we provide both the median response and the fraction of unfavorable responses in figure \ref{fig1:median_unfav}. While there is little discrimination between items when considering the median (or most frequent) score, this figure highlights more unfavorable responses for items related to compensation (“I am satisfied with my compensation”, “I believe my compensation is in line with my experience”), wellbeing (“I can disconnect from my work”) and misconduct (“I can report misconduct without fear of retaliation”).

\subsubsection{Question categories} 
\begin{figure}[h!]
    \centering
    \includegraphics[width=0.95\linewidth]{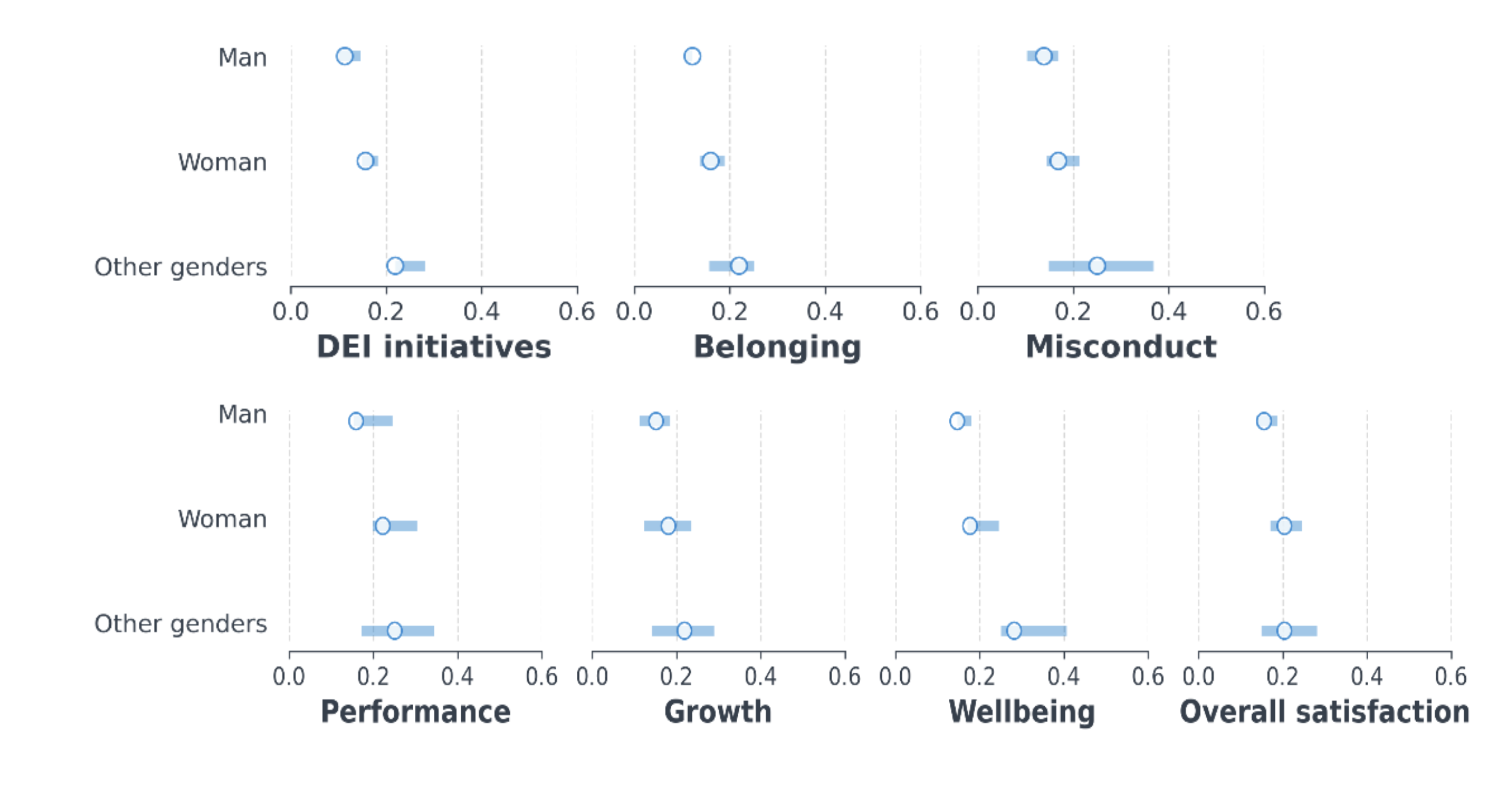} 
    \caption{Fraction of unfavourable response by gender across all question categories.}
    \label{fig1:all-question-cat}
    \end{figure}

When investigating questions, we notice that the fraction of unfavorable responses stems more from certain categories, as shown in figure \ref{fig1:all-question-cat}. Some are consistent across demographic groups, such as misconduct and wellbeing. We also note that participants with a disability or a chronic condition report less satisfaction with how their performance is assessed (Student’s t test on ‘My performance is evaluated fairly’: $p=0.0077$).

\subsubsection{Other factors of variation} 
\begin{figure}[h!]
    \centering
    \includegraphics[width=0.95\linewidth]{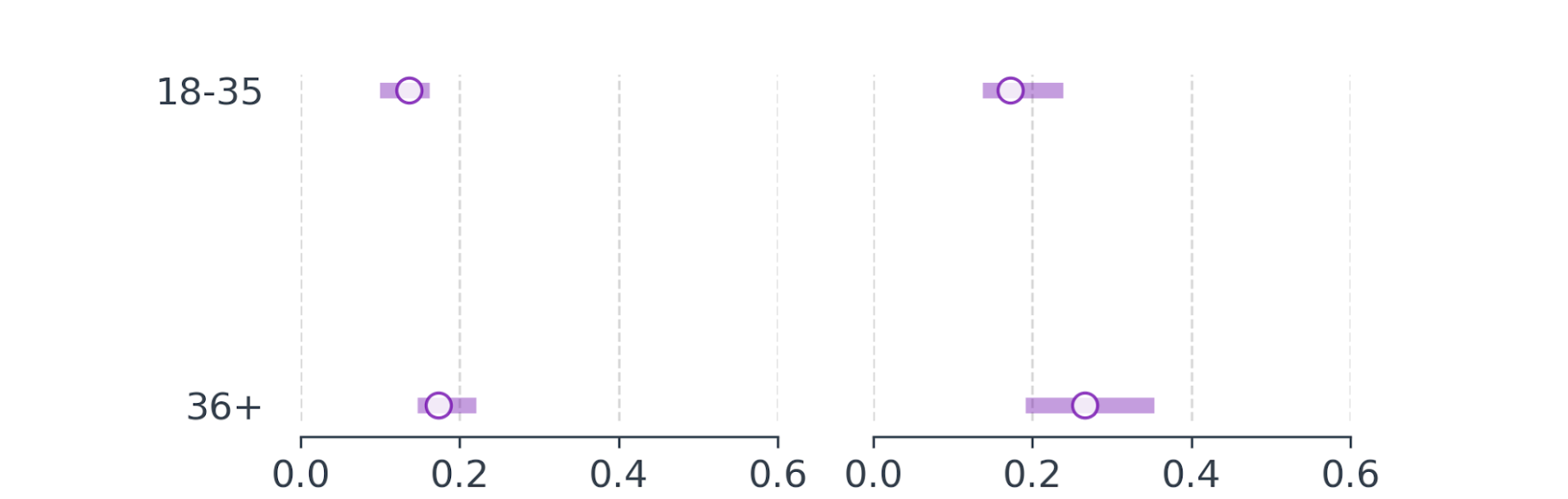} 
    \caption{Fraction of unfavourable response by age across gender majority and minority.}
    \label{fig1:all-question-age-cat}
    \end{figure}

We also investigate other factors of variation where possible. For instance, we observe similar patterns of unfavorable responses between genders in both academia and industry. However, we notice that the level of seniority affects this pattern. While we do not have enough participants identifying as mid-career or senior men, we use binarized age as a proxy and see an effect of age on gender minorities, and to a lesser extent, gender majority (figure \ref{fig1:all-question-age-cat}).

\subsection{Open text comments from participants}
77 participants provided a written comment in an open text box accompanying the final 'Submit' button. They were asked if they had 'Any further feedback on the survey itself, or suggestions?' Out of the responses, 28 were 'none' or 'NA', likely suggesting that participants thought this question required an answer (which it did not). 13 responses provided positive feedback on the survey, including 'thank you' and 'Great survey. Make us think about the dynamics that we usually see with our employers'. 26 respondents provided feedback on the survey itself, suggesting less entanglement in the race and ethnicity demographic question, pointing to the lack of a North African option, suggesting the inclusion of religion or allowing for an option between 'never' or 'multiple times a year'. 2 respondents mentioned that the survey seemed 'oriented' or 'biased', but did not elaborate in the components that made them feel that way. Finally, 9 respondents provided further detail on their personal experiences, with 3 respondents mentioning retaliation for reporting misconduct or sexual harassment and 2 respondents describing instances of 'benevolent sexism' in which opportunities such as teaching or organizing conferences is taken away to 'over-correct' for evidence of unfair administrative and service work. These reports illustrate a willingness to engage further in understanding workplace experiences, and highlight qualitative work, e.g. via interviews, as an important future work.

\subsection{Data analysis} \label{asec:data_analysis}
The completed responses were exported from SurveyMonkey as a spreadsheet to allow for the manual cleaning of some Employer names, as participants were presented with a list of 123 academic or 100 industry employers, or could add their own. No other data was manually revised or filtered.

The spreadsheet was then analyzed in a Python notebook with pandas~\cite {mckinney2010data}, NumPy~\cite{svensson2001construction}, matplotlib~\cite{hunter2007matplotlib}, and scipy~\cite{virtanen2020scipy} to aggregate and plot the responses, compute descriptive statistics, and perform basic statistical testing.

\section{Full Survey}\label{asec:full_survey}

The following questions were included in the survey. Each section of questions (excluding the Microaggression questions) were answerable via a four level Likert scale: Strongly agree, agree, disagree, strongly disagree. Participants were also able to respond “NA/I don’t know” to each question. The Microaggression questions focused on frequency of experiences, and were rated as how often participants experienced the microaggression ranging from daily to a few times per year. Participants could also select never for any microaggression. Microaggression and Accessibility questions were opt in only. 

\subsection{DEI Initiatives} \label{asubsec:DEI_initiatives}
These are regularly mentioned and advertised by Employers. Help us understand how you believe your Employer approaches them.

Suggested: 4 points Likert-scale (strongly agree to strongly disagree) and NA/I don’t know

\begin{enumerate}
    \item My employer states that they value DEI (whether or not they implement it)
    \item My employer demonstrates a visible commitment to diversity and inclusion
    \item My employer provides programs to support underrepresented groups in ML, ex. ERGs, mentoring initiatives, celebration and events
    \item Leaders and employees from majority groups participate in DEI programs and events
    \item I am supported to engage in diversity and inclusion initiatives with no penalty, or a positive impact on my work performance.
    \item DEI programs are initiated by:

    \begin{itemize}
        \item Members of underrepresented groups
        \item Internal committees 
        \item Senior leadership
        \item No DEI programs
    \end{itemize}
\end{enumerate}

\subsection{Belonging} \label{asubsec:belonging}
Employees who feel a sense of belonging at their organisation engage more with their work and Employer. Please help us understand your experience.

Suggested: 4 points Likert-scale (strongly agree to strongly disagree) and NA/I don’t know

\begin{enumerate}
    \item My perspective is valued, even when it is different from others
    \item I am treated with respect
    \item I can be myself at work
    \item I feel safe bringing up difficult subjects or reporting mistakes
    \item I feel part of a workplace community
\end{enumerate}

\subsection{Accessibility} \label{asubsec:accessibility}
People with a disability sometimes need accommodations to access and perform well in the workplace. Please assess your Employer’s accessibility initiatives if you are in this situation.

\begin{itemize}
    \item \textbf{Optional} Do you identify as a person with a disability or other chronic condition?
    \begin{itemize}
        \item Prefer not to say
        \item Yes
        \item No
    \end{itemize}

\end{itemize}

\begin{enumerate}
    \item I feel comfortable discussing physical or cognitive challenges
    \item My employer accommodates my disability
    \item My employer provides different options to support me in my role given my disability, e.g., flexible work arrangements and/or material
    \item I am confident that requesting accommodations will not harm my career
\end{enumerate}

\subsection{Microaggressions} \label{asubsec:microaggressions}

Now we'd like you to think about incidents that might be unpleasant in the workplace, specifically microaggressions. Please help us understand your experiences.

\begin{itemize}
    \item I experience microaggressions because I am part of an underrepresented group. 
    \item Frequency options:
    \begin{itemize}
        \item Prefer not to say
        \item Never
        \item Multiple times a year
        \item Multiple times a month
        \item Multiple times a week
        \item Multiple times a day
    \end{itemize}
\end{itemize}

\subsubsection{Follow-up Question (Conditioned on response)}
\begin{itemize}
    \item Are you willing to describe your experience in the following questions? Please be aware that this can trigger negative reactions if you have experienced microaggressions.
    \begin{itemize}
        \item Yes
        \item No
    \end{itemize}
\end{itemize}

\subsubsection{Trigger Warning: Discussion of Microaggressions}

Please rate how often you experience the following microaggressions in the workplace.\\

\textbf{Response scale (suggested table columns):}
\begin{itemize}
    \item Prefer not to say
    \item Never
    \item Multiple times a year
    \item Multiple times a month
    \item Multiple times a week
    \item Multiple times a day
    \item NA / I don't know
\end{itemize}

\subsection{Misconduct} \label{asubsec:misconduct}
Sometimes, things go wrong. Help us understand how your Employer addresses these situations.
 
Suggested: 4 points Likert-scale (strongly agree to strongly disagree) and NA/I don’t know

\begin{enumerate}
    \item My employer provides ways of reporting misconduct and ensures all employees are aware of it
    \item I can report misconduct without fear of retaliation
    \item My employer responds quickly and consistently to reports of misconduct
    \item If you have reported misconduct previously: I am satisfied with how my report has been handled
\end{enumerate}

\subsection{Performance and Compensation} \label{asubsec:performance_compensation}
Help us understand whether you are satisfied with the way your Employer evaluates your performance and determines your compensation.

Suggested: 4 points Likert-scale (strongly agree to strongly disagree) and NA/I don’t know

\begin{enumerate}
    \item I can succeed to my full ability
    \item I am asked to perform service or volunteer work at the same frequency as my colleagues
    \item My performance is evaluated fairly
    \item When I do an excellent job, my achievements are rewarded
    \item I am satisfied with my compensation
    \item I believe my compensation is in line with my experience
    \item I am satisfied with the type of contract (full time, part time, temporary) I have
\end{enumerate}

\subsection{Growth} \label{asubsec:growth}
Please tell us about growth opportunities at your Employer.

Suggested: 4 points Likert-scale (strongly agree to strongly disagree) and NA/I don’t know

\begin{enumerate}
    \item I am continually learning and growing in my role
    \item My employer provides opportunities for continued growth and development
    \item These opportunities are provided to all employees equitably
    \item There are mentorship programs, or other support, to help me grow my career
\end{enumerate}

\subsection{Well Being} \label{asubsec:well_being}
While a good workplace is important, it’s also important to be able to disconnect from work. In this regard, please tell us about your work-life balance.

Suggested: 4 points Likert-scale (strongly agree to strongly disagree) and NA/I don’t know

\begin{enumerate}
    \item My employer provides great options for leave and time-off, e.g., exceeding legal requirements in my country
    \item My employer supports me in difficult times (e.g., COVID measures, exceptional personal situations)
    \item My employer cares about my well-being
    \item I can disconnect from my work
    \item I am satisfied with my overall well-being
\end{enumerate}

\subsection{Overall Satisfaction} \label{asubsec:overall_satisfaction}
Finally, at a high-level, please tell us about your overall satisfaction with your Employer.

Suggested: 4 points Likert-scale (strongly agree to strongly disagree) and NA/I don’t know

\begin{enumerate}
    \item My career goals can be met at my employer
    \item I am satisfied by the level of agency my employer provides me
    \item My employer acts on the feedback provided by its employees
    \item I would recommend my employer as an excellent place to work
\end{enumerate}